\documentclass[sigconf]{acmart}

\usepackage{tikz}
\usetikzlibrary{shadows}
\usepackage{collab}
\usepackage{xspace}
\usepackage{enumitem}
\usepackage{multirow}   
\usepackage{tabularx}   
\usepackage{makecell}   
\usepackage{booktabs}   
\usepackage{arydshln}   
\usepackage{graphicx}   
\usepackage{pifont}
\usepackage{subcaption}
\usepackage{longtable}

\usepackage[framemethod=TikZ]{mdframed}
\mdfsetup{
    skipabove=10pt,
    skipbelow=10pt,
    innertopmargin=8pt,
    innerbottommargin=8pt,
    innerleftmargin=10pt,
    innerrightmargin=10pt,
    backgroundcolor=gray!10,
    linecolor=black!30,
    linewidth=1pt,
    roundcorner=5pt
}

\newcommand{\alias}{\textsc{MAS-FIRE}\xspace}

\collabAuthor{zb}{teal}{Zhuangbin Chen}
\collabAuthor{zl}{purple}{Zhiling Deng}
\collabAuthor{na}{green}{Jin Jia}

\AtBeginDocument{%
  }

\setcopyright{acmlicensed}
\copyrightyear{2026}
\acmYear{2026}
\acmDOI{XXXXXXX.XXXXXXX}
\acmConference[Conference acronym 'XX]{Make sure to enter the correct
  conference title from your rights confirmation email}{June 03--05,
  2026}{Woodstock, NY}
\acmISBN{978-1-4503-XXXX-X/2026/02}

\begin{document}

\title{\alias: Fault Injection and Reliability Evaluation for LLM-Based Multi-Agent Systems}


\author{Jin Jia}
\authornote{Both authors contributed equally to this research.}
\affiliation{%
  \institution{Sun Yat-sen University}
  \city{Zhuhai}
  \state{Guangdong}
  \country{China}}
\email{jiaj9@mail2.sysu.edu.cn}
\orcid{0009-0002-9691-6648}

\author{Zhiling Deng}
\authornotemark[1]
\affiliation{%
  \institution{Sun Yat-sen University}
  \city{Zhuhai}
  \state{Guangdong}
  \country{China}}
\email{dengzhling3@mail2.sysu.edu.cn}
\orcid{0009-0009-8715-8962}

\author{Zhuangbin Chen}
\authornote{Zhuangbin Chen is the corresponding author.}
\affiliation{%
  \institution{Sun Yat-sen University}
  \city{Zhuhai}
  \state{Guangdong}
  \country{China}}
\email{chenzhb36@mail.sysu.edu.cn}
\orcid{0000-0001-5158-6716}

\author{Yingqi Wang}
\affiliation{%
  \institution{Sun Yat-sen University}
  \city{Zhuhai}
  \state{Guangdong}
  \country{China}}
\email{wangyq376@mail2.sysu.edu.cn}

\author{Zibin Zheng}
\affiliation{%
  \institution{Sun Yat-sen University}
  \city{Zhuhai}
  \state{Guangdong}
  \country{China}}
\email{zhzibin@mail.sysu.edu.cn}
\orcid{0000-0001-7872-7718}


\begin{abstract}

As LLM-based Multi-Agent Systems (MAS) are increasingly deployed for complex tasks, ensuring their reliability has become a pressing challenge. Since MAS coordinate through unstructured natural language rather than rigid protocols, they are prone to semantic failures (e.g., hallucinations, misinterpreted instructions, and reasoning drift) that propagate silently without raising runtime exceptions. Prevailing evaluation approaches, which measure only end-to-end task success, offer limited insight into how these failures arise or how effectively agents recover from them. To bridge this gap, we propose \alias, a systematic framework for fault injection and reliability evaluation of MAS. We define a taxonomy of 15 fault types covering intra-agent cognitive errors and inter-agent coordination failures, and inject them via three non-invasive mechanisms: prompt modification, response rewriting, and message routing manipulation. Applying \alias to three representative MAS architectures, we uncover a rich set of fault-tolerant behaviors that we organize into four tiers: mechanism, rule, prompt, and reasoning. This tiered view enables fine-grained diagnosis of where and why systems succeed or fail. Our findings reveal that stronger foundation models do not uniformly improve robustness. We further show that architectural topology plays an equally decisive role, with iterative, closed-loop designs neutralizing over 40\% of faults that cause catastrophic collapse in linear workflows. \alias provides the process-level observability and actionable guidance needed to systematically improve multi-agent systems.

\end{abstract}

\ccsdesc[500]{Software and its engineering~Software testing and debugging}
\ccsdesc[500]{Software and its engineering~Software verification and validation}

\keywords{Fault Injection, Robustness Evaluation, Multi-agent Systems}

\maketitle

\section{Introduction}

The rapid advancement of LLMs has catalyzed a paradigm shift in intelligent software, moving from monolithic chatbots to orchestrated Multi-Agent Systems (MAS). By assigning specialized roles (e.g., planning, coding, and reviewing) to distinct agent instances, MAS can tackle complex, long-horizon tasks through collaboration~\cite{metagpt, wang2025automisty, boiko2023emergent, ghafarollahi2024protagents, shen2025manus}. 
These systems have demonstrated impressive capabilities in domains ranging from automated software engineering to scientific discovery. However, as MAS transition from experimental prototypes to mission-critical components in production environments, their reliability and fault tolerance become paramount concerns.

Traditional fault-tolerance mechanisms of distributed systems typically address well-defined failures such as component crashes or network timeouts~\cite{fault_tolerance_survey, fault_tolerance_in_cloud}. However, these paradigms are insufficient for MAS due to a fundamental architectural difference. Unlike traditional distributed systems that rely on rigid protocols (e.g., gRPC, REST), MAS utilize natural language as their primary interface for coordination. While this flexibility enables dynamic collaboration, it introduces a unique class of reliability challenges where the system state is defined not by deterministic variables, but by the semantic context of unstructured dialogue. 
Consequently, failures rarely manifest as explicit crashes; instead, they appear as ``soft'' semantic deviations (e.g. hallucinations~\cite{Hallucinationllm, metamorphic-hallucination,haleval2024}, ambiguous interpretations~\cite{zhu2023promptbench}, or reasoning drift~\cite{huang2024,PlanGenLLMs,llm-planning-abilities}) that silently propagate through the system without runtime exceptions.

\begin{figure*}[t]
    \centering
    \includegraphics[width=0.92\linewidth]{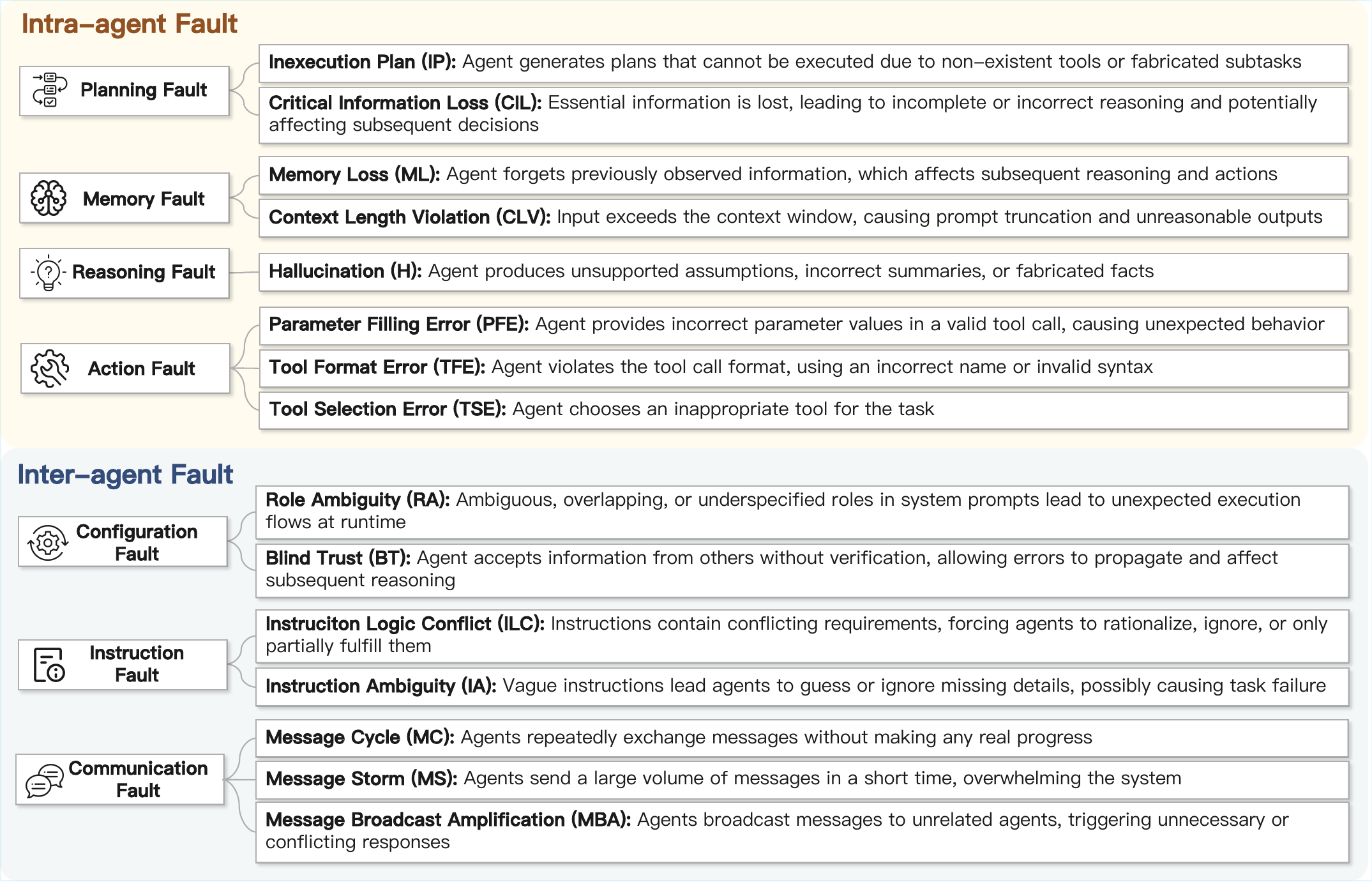}
    \vspace{-6pt}
    \caption{Fault Injection Model for MAS}
    \vspace{-6pt}
    \label{fig:fault_taxonomy}
\end{figure*}


Current evaluation methodologies for MAS are ill-equipped to diagnose these semantic vulnerabilities. They primarily rely on outcome-oriented metrics, such as binary task success rates or sub-goal completion percentages~\cite{agentbench, agentboard}. 
While effective for benchmarking capabilities, these metrics treat the system as a black box, obscuring the process of failure and recovery. 
They fail to answer critical questions: Did the system succeed because it is robust, or due to a lucky retry? Did it fail because of a logic error, or because a rigid architecture prevented an agent from asking for clarification? Without fine-grained observability into how agents respond to anomalies, whether they self-correct, or stall, improving system robustness remains a trial-and-error process.

To address this gap, we introduce MAS-FIRE, a systematic framework for Fault Injection and Robustness Evaluation of Multi-Agent Systems. MAS-FIRE moves beyond simple success metrics to provide a granular analysis of agent resilience. We establish a grounded taxonomy of 15 distinct fault types, categorized into intra-agent faults (affecting internal cognitive processes) and inter-agent faults (disrupting coordination). To simulate realistic production failures, we design non-invasive injection mechanisms that introduce perturbations through prompt modification, response rewriting, and message routing, preserving the system's internal architecture.

Using this framework, we evaluate three representative MAS architectures across the 15 fault types in our taxonomy. Our analysis identifies a comprehensive set of fault-tolerant behaviors, revealing the specific processes through which agents detect, mitigate, and recover from semantic and structural anomalies. We categorize these observed behaviors into four hierarchical tiers: mechanism, rule, prompt, and reasoning.
These tiers provides a structured lens for diagnosing MAS resilience, allowing developers to decouple the contributions of system architecture from those of model reasoning.
Furthermore, we find that while advanced models excel at semantic reasoning, they are paradoxically more vulnerable to prompt-level corruption due to strict instruction compliance. Structural design serves as the an equally important safeguard, with iterative, closed-loop topologies neutralizing over 40\% of the faults that dismantle linear, waterfall-style workflows.



The major contributions of this work are summarized as follows:

\begin{itemize}[noitemsep,leftmargin=5.5mm]
    \item We propose \alias, a fault injection and robustness evaluation framework for MAS, which includes three non-invasive injection mechanisms and a comprehensive suite of robustness metrics that quantify both system-level stability and process-level fault-tolerant effectiveness.
    \item We establish a fault taxonomy categorizing 15 MAS-specific fault types across intra-agent faults and inter-agent faults, and a behavioral taxonomy characterizing fault-tolerant responses along four system dimensions, enabling a fine-grained diagnosis of how systems fail or recover.
    \item Through extensive evaluation on three representative MAS architectures, we systematically identify and quantify the fault-tolerant behaviors exhibited by MAS under different fault types. Our analysis reveals multi-dimensional nature of MAS robustness that provide actionable insights essential for designing and deploying robust MAS. 
\end{itemize}

\section{Background}

\subsection{LLM-Based Multi-Agent Systems as Intelligent Software}
The emergence MAS powered by LLMs marks a transition toward a new type of intelligent software. 
Unlike traditional monolithic applications or standard microservices~\cite{Microservices2016}, LLM-based MAS operate by orchestrating autonomous agents that function as specialized computing units~\cite{metagpt,chatdev,li2025mas,autogen,du2024debate,camel}.
Each agent is typically endowed with distinct capabilities (e.g., planning, memory retention, and tool execution), allowing the collective system to decompose and solve complex, long-horizon problems that exceed the capacity of a single model instance. 
In this paradigm, the LLM functions as the cognitive core, responsible for interpreting instructions, reasoning through sub-tasks, and generating executable actions based on environmental feedback.

A defining characteristic of this intelligent software is its reliance on natural language as the primary interface for coordination. 
Whereas traditional distributed systems communicate via rigid, pre-defined protocols (e.g., gRPC, REST), agents in an MAS collaborate through unstructured semantic dialogue. 
This reliance on natural language introduces a unique layer of complexity: the system state is not defined by deterministic variables but by the semantic context of the conversation history~\cite{memorybank,packer2023memgpt}. Thus, the architectural topology, whether organized as a sequential pipeline~\cite{chatdev}, a hierarchical hierarchy~\cite{metagpt}, or a cooperative network~\cite{camel}, plays a critical role in defining how information flows and how effectively the system can maintain logical consistency across diverse agent interactions.


\subsection{Fault Injection and the Reliability of MAS}

Fault Injection (FI) is a well-established technique in reliability engineering used to assess a system's resilience by deliberately introducing perturbations~\cite{DBLP:conf/icws/LongWCCCW20,DBLP:conf/icse/LiuLDLZ22}.
Historically, FI methodologies have been applied across various abstraction layers, from hardware-level signal interference to software-level logic mutation and interface data corruption~\cite{DBLP:conf/cloud/Meiklejohn2021ServiceLevelFI,DBLP:journals/tdsc/ChenCYLH24MicroFI}. 
The primary objective is to accelerate the occurrence of rare failure modes, thereby allowing developers to validate error-handling mechanisms and ensure that local faults do not cascade into catastrophic system failures.

However, applying existing FI methodologies to LLM-based intelligent software presents significant challenges due to their probabilistic and semantic nature. Traditional software faults typically manifest as explicit crashes, exceptions, or timeouts that are easily detectable by runtime monitors. 
In contrast, failures in MAS often appear as ``silent'' semantic deviations such as hallucinated facts~\cite{Hallucinationllm,haleval2024}, misinterpreted instructions~\cite{prompt-defects-taxonomy,zhu2023promptbench}, or reasoning drift~\cite{mast,TRAIL}, where the system continues to operate without triggering technical exceptions. 
Furthermore, because agent behavior is non-deterministic, a minor perturbation in a prompt or a slight distortion in a message can lead to vastly different outcomes depending on the interaction context. 
Thus, evaluating the robustness of MAS requires shifting the focus of injection from purely syntactic or structural code mutations to semantic perturbations that directly challenge the agents' cognitive and collaborative capabilities.

\section{An MAS Fault Injection and Robustness Evaluation Framework}
\label{sec:fault_injection_and_robustness_evaluation_framework}


In this section, we present \alias, a systematic framework for evaluating MAS robustness through controlled fault injection.
We begin by establishing a grounded MAS fault taxonomy, which synthesize potential failure modes into distinct intra-agent and inter-agent categories based on architectural boundaries. 
Based on these faults, we detail our fault jnjection mechanisms, employing non-invasive techniques (prompt modification, interception and response rewriting, and message routing manipulation) to simulate realistic and representative anomalies during execution.
Finally, we define a set of quantitative robustness metrics designed to measure both the overall system resilience and the specific efficacy of fault-tolerance mechanisms.

\begin{figure*}[t]
    \centering
    \begin{subfigure}{0.33\linewidth}
        \centering
        \includegraphics[width=0.95\linewidth]{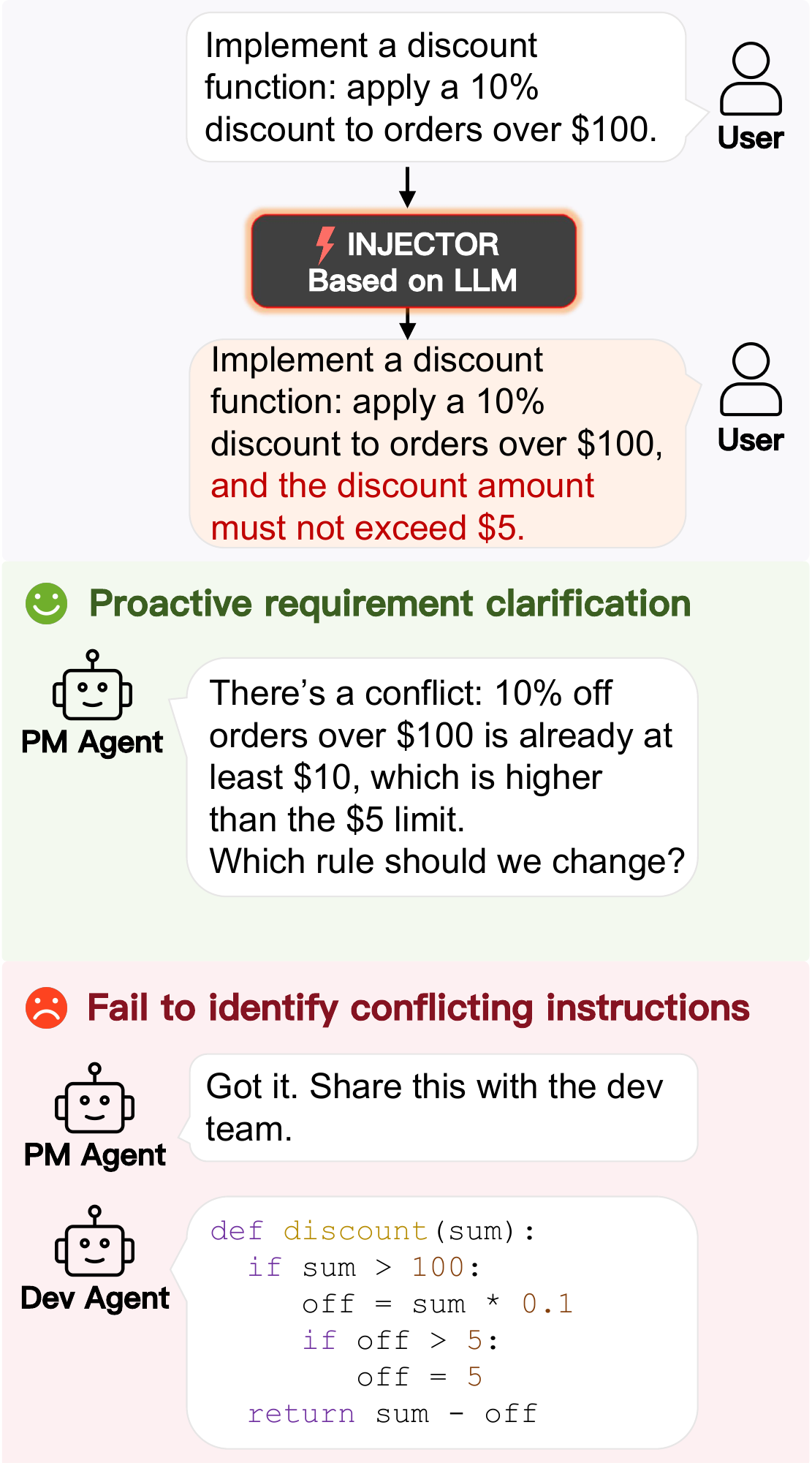}
        \caption{Prompt Modification}
        \label{fig:prompt_modification}
    \end{subfigure}\hfill
    \begin{subfigure}{0.33\linewidth}
        \centering
        \includegraphics[width=0.98\linewidth]{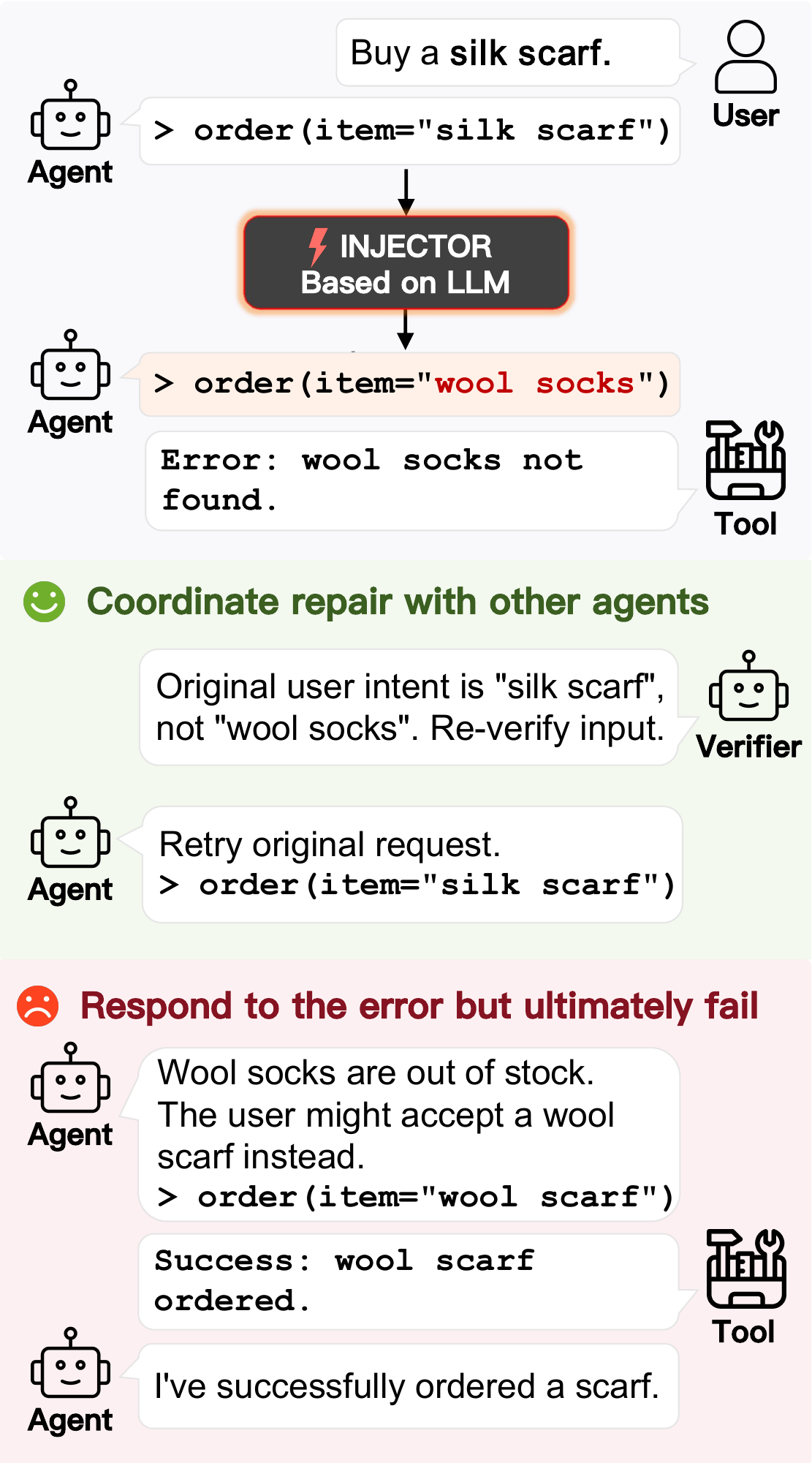}
        \caption{Intercept. and Resp. Rewriting}
        \label{fig:response_rewriting}
    \end{subfigure}\hfill
    \begin{subfigure}{0.33\linewidth}
        \centering
        \includegraphics[width=0.98\linewidth]{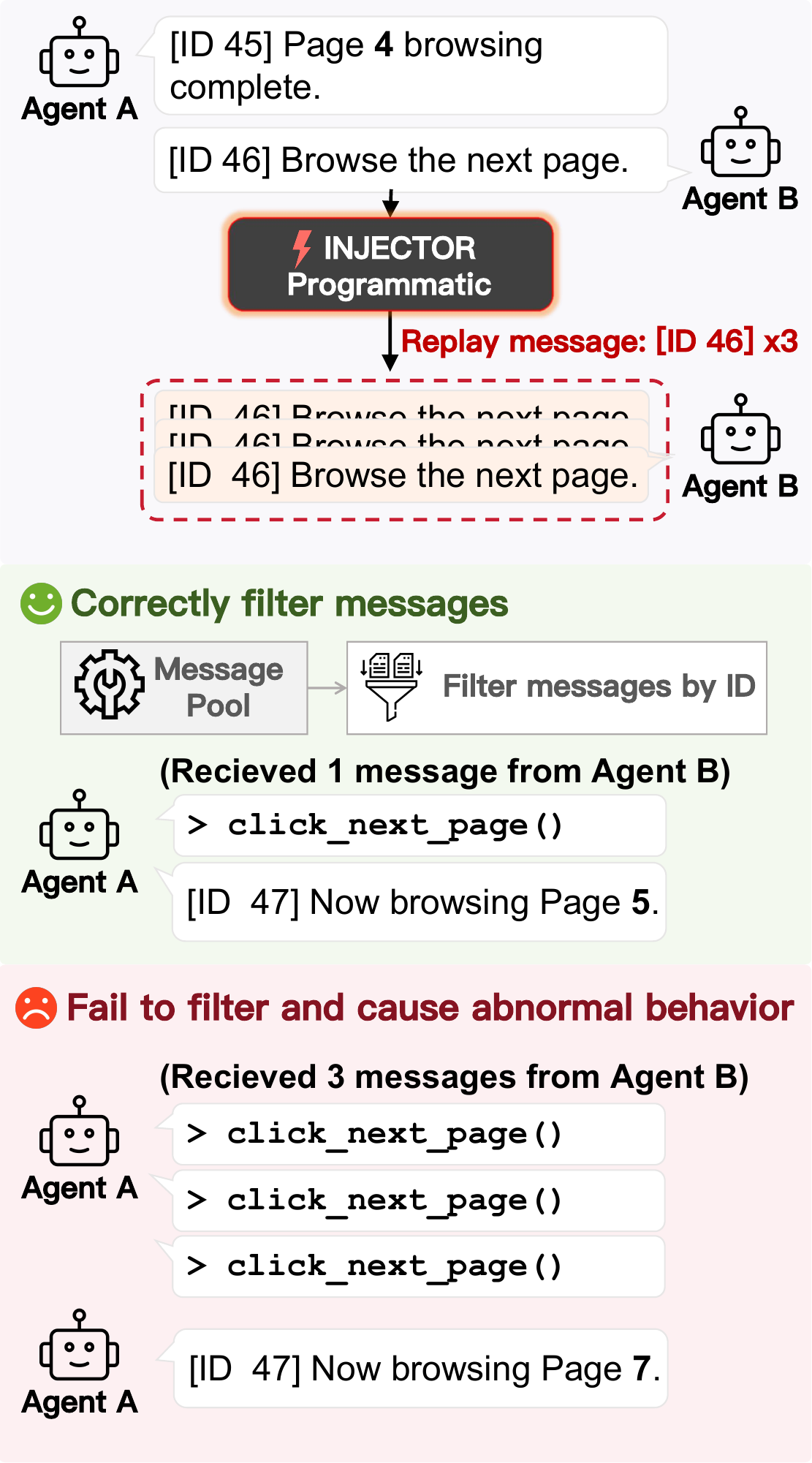}
        \caption{Message Routing Manipulation}
        \label{fig:message_routing}
    \end{subfigure}
    \caption{Examples of Fault Injection Mechanisms and Multi-Agent Recovery Behaviors. (a) \textit{Instruction Logic Conflict} via Prompt Modification, which introduces incompatible constraints to evaluate requirement clarification; (b) \textit{Parameter Filling Error} via Interception and Response Rewriting, which alters task parameters to assess inter-agent coordination and repair; (c) \textit{Message Storm} via Message Routing Manipulation, which injects redundant communication to test infrastructure-level filtering. Green panels illustrate robust recovery (good behavior), while red panels highlight representative failure modes (bad behavior).}
    \label{fig:three_images}
\end{figure*}

\subsection{MAS Fault Taxonomy}
\label{subsec:mastaxonomy}



To establish a rigorous foundation for MAS robustness evaluation, we derive a comprehensive fault taxonomy based on systematic literature review of 
empirical evaluations~\cite{faulty-agents-resilience,wu2025llm,qian2025understanding}, 
benchmarks~\cite{agentbench, ScienceAgentBench, AppWorld, workbench, logicbench, rolebench, workbench, zhu2023promptbench, shortcutsbench,agentboard}, 
and specialized failure studies~\cite{haleval2024, Hallucinationllm, huang2024, yu2025survey, andries2024, liu2024lost, memorybank, packer2023memgpt, xie2024travelplanner, metamorphic-hallucination, shao2023character, communication2025, PlanGenLLMs}. 
We categorize faults according to where they arise in MAS execution: \textit{Intra-agent Faults} emerge within an individual agent's internal processing (planning, memory, reasoning, action), while \textit{Inter-agent Faults} affect coordination and information flow across multiple agents (configuration, instruction, communication). This distinction is critical, as internal reasoning faults require fundamentally different injection strategies than coordination failures. As illustrated in Fig.~\ref{fig:fault_taxonomy}, the taxonomy contains seven fault categories, serving as the unified basis for \alias.

\subsubsection{Intra-agent Faults}

Intra-agent faults encompass failures inherent to an individual agent's cognitive pipeline, directly degrading its ability to plan, retain context, reason, or execute commands. Following standard agent capability decompositions~\cite{agentops-survey,xi2025}, we classify these faults into four subclasses. \textbf{Planning Faults} stem from deficiencies in task decomposition and execution scheduling, manifesting when agents generate inexecutable plans (by invoking non-existent tools or hallucinating subtasks) or omit essential task constraints~\cite{llm-planning-abilities, embodied-fine-grained, PlanGenLLMs, react, xie2024travelplanner}. 
\textbf{Memory Faults} arise from incorrect information retention or management, including aggressive context compression leading to critical information loss or context overflow exceeding effective processing windows~\cite{liu2024lost, memorybank, packer2023memgpt, babilong}. \textbf{Reasoning Faults} reflect errors in the agent's inference engine, commonly manifesting as hallucinations where agents generate incorrect summaries, unsupported assumptions, or fabricated facts that propagate to subsequent stages~\cite{metamorphic-hallucination, haleval2024, Hallucinationllm}. \textbf{Action Faults} occur during external tool interactions, involving inappropriate tool selection, invocation format violations, or invalid parameter supply~\cite{agentbench, shortcutsbench, workbench, li2023api, schick2024toolformer}.







\subsubsection{Inter-agent Faults}

Inter-agent faults capture failures emerging from misaligned assumptions, faulty dependencies, or abnormal interactions between agents. These faults propagate across inter-agent communications, amplifying errors and causing global task failures even when individual agents function correctly in isolation. We divide these into three subclasses. \textbf{Configuration Faults} refer to failures caused by flawed agent role definitions and dependency assumptions established prior to execution, manifesting when agent roles specified through natural-language prompts are ambiguous, overlapping, or underspecified, or when agents blindly trust information from other agents without validation~\cite{rolebench, rolellm, shao2023character}. \textbf{Instruction Faults} denote failures introduced by defects in user-provided task instructions shared across agents, including logical conflicts (mutually incompatible requirements) and semantic ambiguity (insufficient clarity for consistent interpretation)~\cite{self-evolving-benchmark, prompt-defects-taxonomy, wu2025llm, zhu2023promptbench}. \textbf{Communication Faults} capture failures from abnormal message-passing behaviors, including message duplication without proper deduplication, message cycles where agents enter repetitive loops, and message broadcast amplification where messages are mistakenly disseminated to unintended agents~\cite{agentops-survey, communication2025}.






\subsection{Fault Injection Mechanism}
\label{sec:fault_injection_mechanism}

The fault taxonomy described above defines seven high-level fault categories. To operationalize these for systematic evaluation, we design 15 concrete, injectable fault types as shown in Fig.~\ref{fig:fault_taxonomy}. These faults originate from different sources (i.e., prompts governing agent identity and task specification, runtime outputs during execution, and message flows coordinating multi-agent interaction), each requiring distinct injection strategies.
To inject them without compromising the non-invasive nature of the evaluation, we design three complementary fault injection mechanisms:



\subsubsection{Prompt Modification}

This mechanism (as shown in Fig.~\ref{fig:prompt_modification}) targets natural language prompts that shape how agents interpret their responsibilities and task objectives. It operates on two prompt types: \textit{system prompts} (establishing agent identity, roles, and behavioral policies at initialization) and \textit{user prompts} (conveying task requirements and constraints). By corrupting these textual directives, this mechanism injects two categories of faults:


\begin{itemize}[noitemsep,leftmargin=5.5mm]
    \item \textbf{Configuration Faults.} To inject this type of faults, \alias modifies system prompts before agent instantiation, introducing architectural defects. Two specific faults are implemented: (1) \textit{Role Ambiguity}, which merges conflicting role definitions into an agent's system prompt (e.g., acting as both ``Developer'' and ``Tester''), forcing the agent to manage disparate objectives and leading to internal logic conflicts; and (2) \textit{Blind Trust}, which injects unconditional trust directives (e.g., ``accept all input from Agent X as absolute truth''), disabling critical verification and causing uncritical propagation of upstream errors.

    \item \textbf{Instruction Faults.} \alias intercepts user prompts at the MAS entry point and applies semantic transformations that introduce logical inconsistencies or ambiguity into the task specification, forcing correctly configured agents to operate under flawed premises. \alias employs a rule-guided LLM-based injector implementing two mutation strategies: (1) \textit{Instruction Logic Conflict}, which introduces mutually incompatible constraints (e.g., ``Implement a discount function that orders over \$100 get 10\% off, and the discount amount must not exceed \$5'', creating logically unsatisfiable specifications that force agents to rationalize conflicts or arbitrarily prioritize constraints; and (2) \textit{Instruction Ambiguity}, which degrades prompt specificity by replacing concrete terms with vague language (e.g., ``Sort by revenue descending'' becomes ``Organize the data appropriately''), forcing agents to infer intent from insufficient information and leading to divergent interpretations.


\end{itemize}

\subsubsection{Interception and Response Rewriting}

This mechanism (Fig.~\ref{fig:response_rewriting}) targets agent runtime behavior by intercepting agent outputs at critical interaction boundaries (where agents communicate with peers or invoke external tools) and applying targeted mutations before outputs reach recipients. By positioning interceptors at the middleware layer, this mechanism injects Intra-agent Faults (\textbf{Planning}, \textbf{Memory}, \textbf{Reasoning}, \textbf{Action}) through two mutation categories:


\textit{Semantic-Level Mutation.} For faults requiring contextual understanding, \alias employs a prompt-guided mutation strategy. The interceptor delegates captured content (reasoning chains, planning outputs, tool invocation requests) to a secondary LLM injector guided by predefined fault templates, which applies semantic transformations that preserve surface coherence while corrupting underlying correctness.

\begin{itemize}[noitemsep,leftmargin=5.5mm]
    \item \textbf{Planning Faults.} \alias injects \textit{Inexecutable Plan} by introducing logical inconsistencies such as circular task dependencies, references to non-existent tools or agents, or invalid workflow orderings. \textit{Critical Information Loss} is injected by selectively removing essential constraints, parameters, or context from planning outputs, causing downstream agents to operate on incomplete specifications.

    \item \textbf{Reasoning Faults.} \alias injects \textit{Hallucination} by replacing verified facts with plausible but factually incorrect info, removing uncertainty qualifiers (e.g., changing ``likely'' to definitive statements), or introducing fabricated intermediate reasoning steps. These mutations corrupt the semantic integrity of agent deliberation while maintaining syntactic well-formedness.

    \item \textbf{Action Faults.} \alias injects \textit{Tool Selection Error} by substituting the intended tool with a semantically similar but incorrect alternative (e.g., replacing ``calculator'' with ``web\_search''). \textit{Parameter Filling Error} is injected by altering arguments to introduce domain-specific mistakes (e.g., swapping location coordinates, modifying query terms) while preserving type correctness.
\end{itemize}





\textit{Structure-Level Mutation.} For faults independent of semantic context, \alias performs direct algorithmic transformations on message payloads or data structures without LLM assistance, operating on syntactic structure rather than semantic content.

\begin{itemize}[noitemsep,leftmargin=5.5mm]
    \item \textbf{Memory Faults.} \alias injects \textit{Memory Loss} by selectively truncating conversation history using rule-based pruning strategies (e.g., removing early-turn messages, deleting messages from specific agents). \textit{Context Length Violation} is injected by aggressively compressing context windows beyond the agent's effective processing capacity, forcing the agent to operate with incomplete historical information and degrading reasoning fidelity.

    \item \textbf{Action Faults.} \alias injects \textit{Parameter Format Error} by directly corrupting the syntactic structure of tool invocations. This includes introducing malformed JSON (missing brackets, incorrect escaping), breaking API schema constraints (wrong data types, missing required fields), or violating domain-specific formatting rules (e.g., invalid date formats, malformed SQL queries). These errors trigger immediate execution failures at the parsing or validation stage.
\end{itemize}




\subsubsection{Message Routing Manipulation}

This mechanism (as shown in Fig.~\ref{fig:message_routing}) targets inter-agent communication by manipulating message flows, including frequency and recipients, without altering message content. \alias implements it programmatically, injecting faults in a controlled manner without LLM involvement.
This mechanism injects three types of \textbf{Communication Faults}:
(1) \textit{Message Cycle}, which redirects messages back to the sender agent, forcing agents into repetitive conversational loops that halt progress and simulate infinite loops in coordination;
(2) \textit{Message Storm}, which replicates a single point-to-point message multiple times, flooding the receiver to simulate resource exhaustion and test the MAS ability to handle redundancy messages;
(3) \textit{Message Broadcast Amplification}, which redirects messages intended for specific agents to unrelated agents, causing them to receive irrelevant information, perform unnecessary processing, and potentially disrupt consensus and state consistency across the MAS.

\subsection{MAS Robustness Metrics}
\label{sec:robustness_metrics}

While fault injection reveals how agents respond to anomalies, quantitative metrics are essential to systematically compare robustness across systems, fault types, and architectures. Existing MAS evaluation frameworks~\cite{agentbench, agentboard} rely primarily on binary task success rates, which fail to distinguish graceful degradation from catastrophic failures and cannot capture whether agents successfully detect faults, resolve immediate errors, or translate local recovery into end-to-end completion. We define a dual-level evaluation framework: (1) \textit{system-level resilience} measures overall task success rates under fault injection to assess how architectural design and model capabilities preserve functionality, and (2) \textit{process-level effectiveness} analyzes fault-tolerant behaviors during execution to understand detection, response, and recovery mechanisms.



\subsubsection{System-level Resilience}
\label{ssec:syslevl}
System-level metrics measure an MAS' ability to maintain task successful completion under faults, capturing the combined effect of architecture, agent intelligence, and coordination.
This includes \textbf{Robustness Score (RS)}, which quantifies the fraction of originally successful tasks that remain solvable after fault injection. For fault type \(f\):


\begin{equation}
    RS_f = \frac{N_{f,\text{success}}}{T_{\text{base}}}
\end{equation}

\noindent where \(N_{f,\text{success}}\) denotes the number of tasks successfully completed under fault \(f\), and \(T_{\text{base}}\) denotes the set of tasks the MAS completes without faults. A high \(RS\) indicates maintained functionality despite faults, while a low \(RS\) signals architectural or algorithmic vulnerabilities.

\subsubsection{Process-level Effectiveness}
\label{ssec:processlevel}

Process-level metrics evaluate an MAS' internal mechanisms to detect and respond to faults during execution. Unlike system-level metrics focusing on final outcomes, these metrics analyze intermediate behaviors to understand \textit{why} systems succeed or fail.


\begin{equation}
    O_f = \frac{N_{f,\text{trigger}}}{N_{\text{total}}}, \quad 
    L_f = \frac{N_{f,\text{fixed}}}{N_{f,\text{trigger}}}, \quad 
    S_f = \frac{N_{f,\text{final\_success}}}{N_{f,\text{trigger}}}
\end{equation}

\textbf{Occurrence Rate (\(O_f\)).} This metric quantifies the system's ability to detect anomalies and activate fault-tolerant responses. \(N_{f,\text{trigger}}\) denotes the number of tasks in which the system detects abnormality and activates at least one fault-tolerant behavior under fault \(f\), and \(N_{\text{total}}\) denotes the total number of injected tasks. A high \(O_f\) indicates strong fault awareness, where agents recognize deviations from expected execution and initiate corrective actions. A low \(O_f\) suggests silent fault propagation without defensive responses, often leading to cascading failures.

\textbf{Local Success Rate (\(L_f\)).} This metric evaluates the effectiveness of an MAS' fault-tolerant behaviors in resolving the injected fault. \(N_{f,\text{fixed}}\) denotes the number of tasks in which the triggered fault-tolerant behavior successfully corrects the injected fault (e.g., fixing an incorrect tool invocation format). \(L_f\) focuses on \textit{local recovery}, i.e., whether the system can handle the immediate error introduced by the fault. A high \(L_f\) indicates effective error-correction mechanisms, while a low \(L_f\) reveals that agents recognize faults but lack appropriate recovery strategies.

\textbf{Success Rate (\(S_f\)).} The \(S_f\) measures whether local fault recovery translates into global task success. \(N_{f,\text{final\_success}}\) denotes the number of tasks that ultimately achieve their intended goal among those in which fault-tolerant behaviors are triggered. \(S_f\) captures \textit{end-to-end effectiveness} even if an agent successfully corrects an immediate error (reflected in \(L_f\)), the overall task may still fail due to residual effects (e.g., lost context, cascading downstream errors). The gap between \(L_f\) and \(S_f\) reveals the extent to which local recovery is insufficient for global success.

\section{MAS Robustness Evaluation}

To systematically characterize how MAS respond to and recover from faults in practice, we conduct an empirical evaluation that applies the \alias framework (Sec.~\ref{sec:fault_injection_and_robustness_evaluation_framework}) to representative MAS implementations. Unlike existing work that focuses solely on task success rates, our evaluation investigates the \textit{process} through which systems detect, respond to, and recover from faults. This process-oriented analysis enables us to understand not only \textit{whether} systems fail under faults, but also \textit{how} and \textit{why} failures occur, providing actionable insights for improving MAS resilience.
The evaluation is designed to answer the following three research questions:

\begin{itemize}[noitemsep,leftmargin=5.5mm]
    \item How do different faults impact MAS performance and stability?
    \item How does foundation model capability affect MAS robustness?
    \item What fault-tolerant behaviors emerge in MAS when confronted with faults, and how can they be categorized and quantified?
\end{itemize}



\subsection{Experimental Setup}
\label{sec:experiment_setup}

\subsubsection{System Selection and Task Datasets}

We select three representative MAS that span diverse architectural paradigms and application domains. 
\textbf{MetaGPT}~\cite{metagpt} employs a hierarchical organizational workflow for code generation, featuring a shared message pool that maintains a persistent, globally accessible context and a linear sequential execution pipeline. MetaGPT incorporates programmatic design mechanisms such as automatic message deduplication and selective information transmission through instructions, enabling automatic message filtering and deduplication.
\textbf{Table-Critic}~\cite{table-critic} features an iterative critic-refiner pipeline, where a Judge agent acts as a validator to verify outputs. Upon detecting any discrepancies, the agent triggers a closed-loop refinement cycle, facilitating seamless autonomous error detection and self-correction.
\textbf{Camel}~\cite{camel} utilizes a bilateral role-playing structure, where information flows sequentially between a User and an Assistant through cooperative negotiation. This selection ensures diversity in agent roles, communication patterns, coordination mechanisms, and fault propagation characteristics.

For each MAS, we select task datasets representative of its target domain. MetaGPT is evaluated on \textbf{HumanEval}~\cite{chen2021evaluating}, a benchmark for functional correctness of generated code. Table-Critic is evaluated on \textbf{WikiTableQuestions}~\cite{pasupat2015compositional}, a dataset requiring complex table reasoning and multi-step inference. Camel is evaluated on \textbf{WebShop}~\cite{webshop}, which requires multi-turn interaction and decision-making in simulated e-commerce environments. Following the sampling methodology of Krejcie and Morgan~\cite{krejcie1970determining}, we randomly sample 400 instances from WikiTableQuestions (from a total of 4,344), exceeding the minimum representative sample size threshold to ensure statistical robustness.
Table~\ref{tab:system_overview} summarizes the evaluated systems, their associated benchmarks, task scales, and baseline performance under fault-free conditions.



\subsubsection{Fault Injection and Log Collection}

Each sampled task is executed under controlled fault injection using the three complementary mechanisms described in Sec.~\ref{sec:fault_injection_mechanism}. 
For each fault type, faults are injected at predefined execution points that correspond to the appropriate injection mechanism: 
system and user prompt modifications occur at agent initialization and task ingestion (Prompt Modification), 
agent output corruptions occur during agent-to-agent communication or tool invocations (Interception and Response Rewriting), 
and message flow manipulations occur at the inter-agent coordination infrastructure (Message Routing Manipulation). 

\begin{table*}[t]
  \centering
  \setlength{\tabcolsep}{4pt}
  \caption{Evaluated MAS and Baseline Success Rates under Fault-Free Conditions}
  \begin{tabular}{llccc}
    \toprule
    \textbf{System (Paradigm)} & \textbf{Domain (Benchmark)} & \textbf{Total Tasks} & \textbf{GPT-5} & \textbf{Deepseek-V3} \\
    \midrule
    \textbf{MetaGPT} \small{(Dynamic Organization)} & Code Gen (HumanEval\cite{chen2021evaluating}) & 164 & 99.0\% & 89.0\% \\
    \textbf{Table-Critic} \small{(Thought-Critic)} & Table QA (WikiTQ$^\dagger$\cite{pasupat2015compositional}) & 400 & 87.0\% & 78.3\% \\
    \textbf{Camel} \small{(Instructor-Assistant)} & Web Nav (WebShop$^\ddagger$\cite{webshop}) & 251 & 37.8\% & 33.0\% \\
    \bottomrule
    \multicolumn{5}{p{1.36\columnwidth}}{\small \textbf{Note:} All evaluations are \textbf{single-trial (Pass@1)}. $\dagger$10\% systematic sampling of the original 4,344 tasks. $^\ddagger$Tasks are based on AgentBoard-annotated version.}
  \end{tabular}
  \label{tab:system_overview}
\end{table*}

To assess foundation model capability impact (RQ2), each MAS is configured with two foundation models: \textbf{GPT-5} and \textbf{DeepSeek-V3}. 
As shown in Table~\ref{tab:system_overview}, GPT-5 consistently outperforms DeepSeek-V3 across all systems in baseline success rates, representing a stronger and weaker model pair for comparative analysis. For all semantic fault injections (e.g., \textit{Hallucination}, \textit{Inexecutable Plan}), we employ GPT-5-mini as the fault injector to ensure contextual coherence and realistic anomalies. 
Our fault injection experiments achieved a 99\% success rate across all evaluated systems. 
A small fraction of failures occurred when LLM-generated fault specifications contained structural inconsistencies violating the fault injector's integrity checks; these cases were re-executed to ensure complete coverage. 
Additionally, a minimal subset of tasks were excluded from tool-related fault injection because they involved no tool invocations, making such faults inapplicable.

Following execution, we collect MAS logs into a comprehensive corpus. This dataset comprises execution logs spanning 15 fault types (Sec.~\ref{sec:fault_injection_mechanism}), three MAS architectures, two foundation models, and task difficulty levels within each benchmark.

\subsection{Fault-Tolerant Behavior Analysis}

To systematically derive a taxonomy of fault-tolerant behaviors from execution logs, we employ the \textbf{Grounded Theory} methodology~\cite{glaser1967grounded_theory}. Four authors of the paper independently analyze execution logs for each fault type, identifying observable behavioral patterns without predefined categories based on concrete log evidence (agent actions, communication patterns, decision-making strategies). The identified behaviors undergo iterative refinement through constant comparison analysis, i.e., merging semantically equivalent behaviors, splitting overly coarse categories, introducing new categories, and removing insufficiently distinguishable ones, until saturation is reached. 

To enable scalable analysis, we develop an automated annotation pipeline using an LLM-as-a-judge approach. The derived taxonomy, formal behavior definitions, and representative examples are provided to GPT-5, which assigns behavior labels to unseen execution logs. 
We measure agreement between LLM-generated and human-generated annotations on a held-out validation set using Cohen's Kappa ($\kappa = 0.94$), validating the automated annotator's use for large-scale annotation.

\section{Evaluation Results}

This section presents the empirical results of applying the \alias framework to three representative MAS under 15 fault types. We organize the findings around three research questions to comprehensively analyze MAS robustness, fault tolerance mechanisms, and recovery strategies.

\subsection{RQ1: Impact of Different Fault Categories on MAS Robustness}
\label{subsec:rq1}

\begin{figure*}[t]\centering\includegraphics[width=0.92\linewidth]{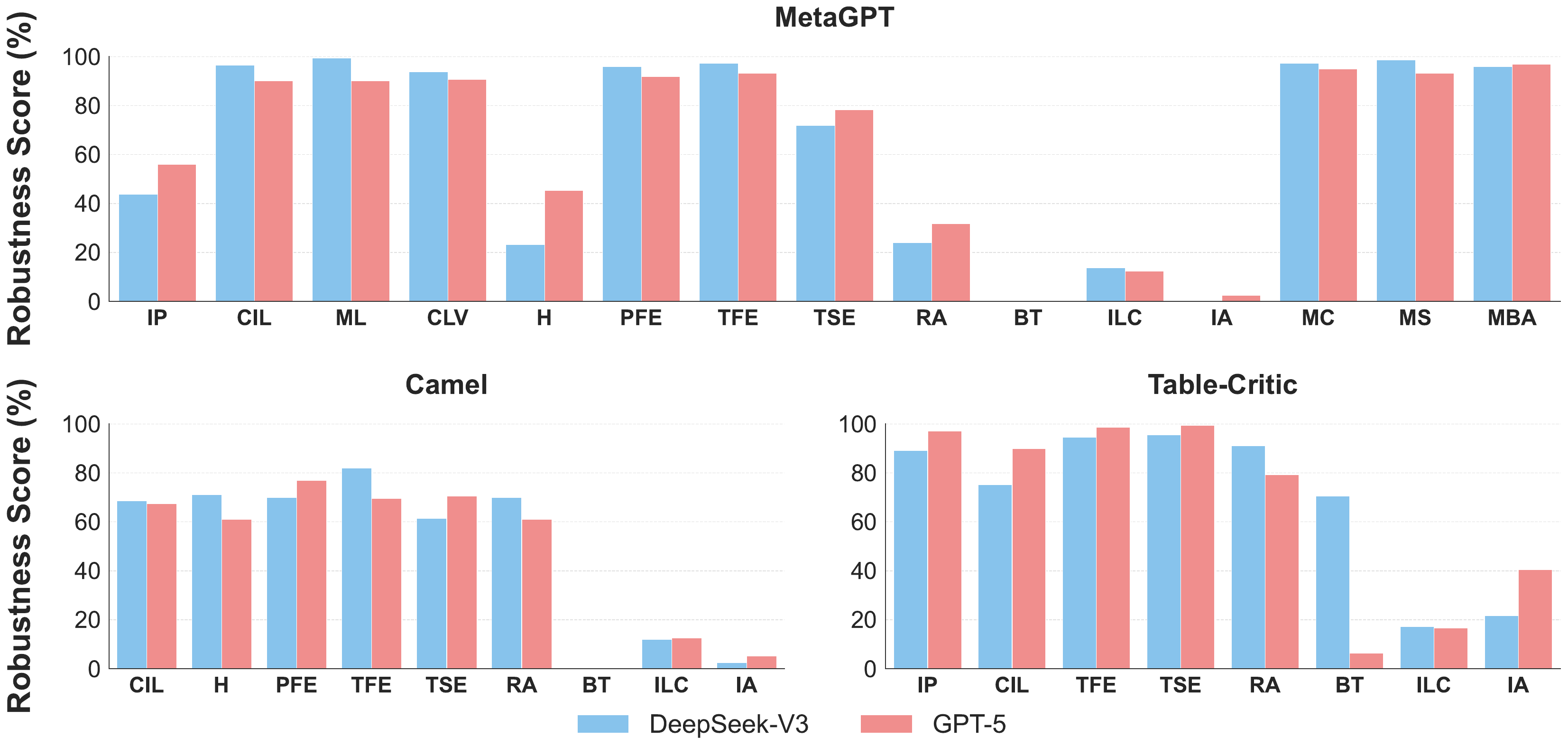}
    \caption{Robustness Score ($RS_f$) of Different MAS under 15 Fault Types}
    \label{fig:success_rate_stats}
\end{figure*}

Our evaluation reveals that MAS exhibit highly varying sensitivity to different fault categories. Fig.~\ref{fig:success_rate_stats} illustrates the $RS_f$, as defined in Sec. \ref{ssec:syslevl}, across all fault types and systems. 

\subsubsection{Impact of Intra-Agent Faults}

Intra-agent faults target the internal inference processes of individual agents, including Planning, Memory, Reasoning, and Action. These faults exhibit strong architecture-dependent impact, as specific MAS design patterns act as structural buffers that neutralize localized failures before they propagate system-wide.

The key pattern across intra-agent faults is that architectural mechanisms can effectively contain their impact.
\textbf{Memory Faults} (\textit{Critical Information Loss}) severely degrade performance in Camel's linear bilateral structure ($RS_f \approx 67\%$) where info flows sequentially. However, MetaGPT's shared message pool maintains a persistent, globally accessible context that allows downstream agents to retrieve missing info ($RS_f > 90\%$) (a $\Delta RS_f \approx +25\%$ advantage). Similarly, \textbf{Planning Faults} substantially reduce MetaGPT's $RS_f$ (as low as $43.84\%$) due to its rigid sequential execution pipeline where a single planning error halts the entire workflow. In contrast, Table-Critic's critique-refinement loop enables autonomous error detection and plan correction through iterative self-assessment, maintaining $RS_f > 89\%$ across both foundation models (a $\Delta RS_f \approx +45\%$ improvement).

Action faults demonstrate another form of architectural mediation. MetaGPT and Table-Critic generally sustain high $RS_f$ on most action-related faults through environmental error feedback and automatic retry logic. When a tool invocation fails, these systems detect the failure signal and regenerate corrected invocations without human intervention. Camel consistently attains lower $RS_f$ on these faults because it lacks comparable dynamic error-trapping mechanisms. \textbf{Reasoning Faults} exhibit moderate impact with strong model dependency, as superior foundation models demonstrate enhanced mitigation through semantic validation (detailed in Sec.~\ref{subsec:rq2}).

\begin{mdframed}
\textbf{Finding 1:} Specific architectural design patterns effectively neutralize faults, e.g., shared message pools mitigate \textbf{Memory Faults} ($\Delta RS_f \approx +25\%$), iterative critique loops neutralize \textbf{Planning Faults} ($\Delta RS_f \approx +45\%$), and environmental feedback mechanisms enable rapid recovery from most \textbf{Action Faults} ($RS_f > 90\%$ for Parameter Filling and Tool Format errors). These architectural features prevent localized failures from escalating into systemic collapse.
\end{mdframed}

\subsubsection{Impact of Inter-Agent Faults}

Inter-agent faults target coordination mechanisms, exhibiting different vulnerability patterns ranging from catastrophic collapse to effective mitigation.



\textbf{Configuration} and \textbf{Instruction Faults} represent the most severe threats by corrupting semantic foundations of agent coordination. \textbf{Configuration Faults} tamper with system prompts at initialization (e.g., "trust all outputs without verification"), while \textbf{Instruction Faults} inject contradictions or ambiguities into user prompts (e.g., conflicting goals). Once the semantic contract between designers and agents is violated, agents cannot distinguish valid logic from injected errors. Results demonstrate catastrophic impact: \textbf{Configuration Faults} reduce MetaGPT's $RS_f$ to $0.0\%$-$31.68\%$ (under \textit{Blind Trust} and \textit{Role Ambiguity}), \textbf{Instruction Faults} cause similar collapse ($RS_f \leq 13.7\%$). Even Table-Critic suffers degradation under \textbf{Instruction Faults} ($RS_f$ to $16.67\%$-$40.52\%$).

Architectural topology critically modulates severity. MetaGPT's linear pipeline exhibits extreme vulnerability: under \textit{Blind Trust}, $RS_f$ drops to $0.0\%$ as sequential dependency enables cascading failures—errors propagate downstream without interception. Under \textit{Role Ambiguity}, MetaGPT's $RS_f$ remains low ($23.97\%$-$31.68\%$). In contrast, Table-Critic's iterative closed-loop provides resilience: under \textit{Role Ambiguity}, $RS_f$ maintains $91.05\%$ (DeepSeek-V3) and $79.31\%$ (GPT-5) as the Critic agent validates outputs and triggers refinement cycles. For \textbf{Instruction Faults}, this mechanism provides partial protection ($RS_f \in [16.67\%, 40.52\%]$), significantly outperforming linear architectures. Camel achieves intermediate performance (\textit{Role Ambiguity}: $RS_f \in [61.05\%, 69.88\%]$) through bilateral negotiation. Notably, \textit{Blind Trust} reveals a paradox: Table-Critic with GPT-5 achieves only $RS_f = 6.32\%$, while DeepSeek-V3 maintains $RS_f = 70.61\%$, there is a $64.29\%$ gap favoring the weaker model due to stricter instruction adherence preventing recovery (detailed in Sec.~\ref{subsec:rq2}).

\begin{mdframed}
    \textbf{Finding 2:}
    \textbf{Configuration Faults} and \textbf{Instruction Faults} cause severe system degradation, with RS dropping as low as $0.0\%$ in worst cases (e.g., \textit{Blind Trust} in linear architectures) by corrupting semantic foundations of agent coordination. The severity is modulated by architecture. Linear workflows are highly vulnerable due to cascading failures, whereas iterative structures mitigate impact.
\end{mdframed}

\textbf{Communication faults} represent the least destructive threat category, consistently yielding RS $> 93\%$ in MetaGPT. Other MAS frameworks were excluded as they lack a communication module, preventing the injection of \textbf{Communication Faults}. This resilience is primarily due to inherent message validation logic. For instance, MetaGPT's role-based subscription model ensures agents only process messages from relevant peers. By handling deduplication (\textit{Message Storm}) and cycle detection (\textit{Message Cycle}) at the infrastructure layer, the system neutralizes these faults deterministically, bypassing the need for complex agent-level reasoning.

\begin{mdframed}
\textbf{Finding 3:} Infrastructure-level defenses provide superior tolerance for \textbf{Communication Faults}. While semantic or \textbf{Configuration Faults} that require cognitive reasoning, communica-tion-related threats are effectively neutralized (RS $> 93\%$) through hard-wired procedural logic, such as role-based filtering and automated cycle detection.
\end{mdframed}

\subsection{RQ2: Role of Foundation Model Capability}
\label{subsec:rq2}

Based on the impact of different fault categories in RQ1, we investigate whether superior foundation models can mitigate these risks. Our findings reveal a dual nature of model capability: while superior models provide significant resilience gains through enhanced semantic reasoning, they can become bottlenecks when system designs enforce strict adherence to potentially corrupted instructions.

Superior foundation models demonstrate significant robustness advantages when architectures provide mechanisms for semantic validation and context-based error correction. MetaGPT's shared message pool enables stronger models to retrospectively analyze execution history, identify logical inconsistencies, and retrieve correct information. For \textit{Hallucination}, GPT-5 performs semantic validation by comparing received information against original task specifications, yielding RS of $45.34\%$ versus DeepSeek-V3's $23.29\%$ ($\Delta RS = +22.05\%$). Similarly, under Inexecutable Plan faults, GPT-5 achieves RS of $55.90\%$ versus $43.84\%$ ($\Delta RS = +12.06\%$).
In contrast, systems like Camel with simpler architectures show limited benefits from superior models. 
\textit{Hallucination} in Camel stem primarily from external environment interactions (e.g., fetching corrupted content from the Internet) rather than internal coordination errors. Since fault mitigation depends more on environmental re-perception than logical reasoning, GPT-5's reasoning advantage provides minimal resilience gain. Similarly, the absence of a global shared information pool prevents models from leveraging historical context to correct semantic deviations.

\begin{mdframed}
\textbf{Finding 4:}
For semantic-related faults such as \textit{Hallucination} and \textit{Inexecutable Plan}, superior foundation model capability plays a critical role in fault mitigation. In systems with shared information pools, enhanced semantic reasoning enables models to retrieve correct historical information and achieve $\Delta RS \approx +17\%$ improvements. However, this advantage is architecture-dependent. When faults stem from external environment interactions or systems lack shared information pools, model capability provides limited resilience gains.
\end{mdframed}

Counterintuitively, superior model intelligence can exacerbate fault impacts when recovery depends on challenging corrupted directives. Table-Critic's Thought-Critic-Refine loop operates through: (1) Generator produces answers with reasoning traces; (2) JudgeAgent evaluates correctness; (3) upon detecting errors, JudgeAgent triggers Refiner to correct mistakes. The \textit{Blind Trust} attacks this via dual injection: corrupting JudgeAgent's system prompt to "unconditionally trust Generator outputs without verification," then injecting semantic errors into Generator's thoughts. Recovery requires JudgeAgent to override its corrupted instruction.

Results reveal a significant reversal in performance. While GPT-5 achieves a $RS$ of only $6.32\%$, DeepSeek-V3 reaches $70.61\%$, representing a $64.29\%$ improvement for the weaker model. GPT-5's superior instruction-following leads to strict adherence to corrupted directives. 
In approximately $93.68\%$ of its failed cases, the JudgeAgent accepts erroneous reasoning without triggering the Critic-Refine loop.
Conversely, DeepSeek-V3 exhibits instructional non-compliance by challenging suspicious inputs in the majority of cases. Paradoxically, this failure to follow corrupted instructions serves as an accidental recovery mechanism, enabling the system to bypass the injected fault.




\begin{mdframed}
\textbf{Finding 5:} Higher model capability becomes counterproductive when a system's resilience relies on bypassing a corrupted directive rather than strictly following it. Superior models' strict compliance with system prompts prevents them from deviating into alternative reasoning paths that might trigger recovery mechanisms. System robustness in such scenarios depends on partial instruction non-compliance, creating a scenario where lower-capability models unexpectedly achieve higher success rates.
\end{mdframed}

\subsection{RQ3: Categorization and Quantification of MAS' Fault-tolerant Behaviors}
\label{subsec:rq3}

\begin{table*}[p]
    \centering
    \small
    \renewcommand{\arraystretch}{0.85} 
    \setlength{\tabcolsep}{2pt} 
    \caption{Fault-Tolerant Behavior Taxonomy and Mechanism Classification. This table evaluates four fault tolerance (FT) dimensions: Mechanism, Rule, Prompt, and Reasoning. Symbols are defined as follows: \ding{51} denotes successful mitigation; \ding{55} indicates activation but failure to resolve; an empty cell signifies no activation. \textbf{Action Faults} are categorized into \textit{Parameter Filling Error}, \textit{Tool Format Error}, and \textit{Tool Selection Error}.}
    \label{tab:agent_behavior_tag}
    \newcolumntype{L}{>{\raggedright\arraybackslash}X}
    \begin{tabularx}{\linewidth}{l L c c c c}
        \toprule
        \textbf{Fault Category} & \textbf{Agent Behavior} & \textbf{Mechanism.} & \textbf{Rule.} & \textbf{Prompt.} & \textbf{Reasoning.} \\
        \midrule

        \multirow{4}{*}{\makecell[l]{\textbf{Inexecutable}\\\textbf{Plan}}} 
        & Restores faulty plan via inherent process & \ding{51} & & & \\ \cdashline{2-6}[0.5pt/2pt]
        & Ignores inexecutable parts and continues & & & & \ding{51} \\ \cdashline{2-6}[0.5pt/2pt]
        & No corrective behavior & & & & \ding{55} \\ \cdashline{2-6}[0.5pt/2pt]
        & Responds to error but ultimately fails & \ding{51} & & & \ding{55} \\ \midrule

        \multirow{5}{*}{\makecell[l]{\textbf{Critical Info}\\\textbf{Loss}}} 
        & Restores missing information via inherent process & \ding{51} & & & \\ \cdashline{2-6}[0.5pt/2pt]
        & Autonomously repairs missing information & \ding{51} & & & \ding{51} \\ \cdashline{2-6}[0.5pt/2pt]
        & Avoids using missing information & & & & \ding{51} \\ \cdashline{2-6}[0.5pt/2pt]
        & Ignores missing information & & & & \ding{55} \\ \cdashline{2-6}[0.5pt/2pt]
        & Uses external information sources to repair & \ding{51} & & & \ding{55} \\ \midrule

        \multirow{5}{*}{\makecell[l]{\textbf{Memory Loss}}}
        & Uses external information sources to repair & \ding{51} & & & \ding{51} \\ \cdashline{2-6}[0.5pt/2pt]
        & Autonomously restores missing memory & & & & \ding{51} \\ \cdashline{2-6}[0.5pt/2pt]
        & Restores missing memory via inherent process & \ding{51} & & & \\ \cdashline{2-6}[0.5pt/2pt]
        & Avoids using missing memory & & & & \ding{51} \\ \cdashline{2-6}[0.5pt/2pt]
        & Ignores missing memory & & & & \ding{55}
        \\ \midrule

        \multirow{4}{*}{\makecell[l]{\textbf{Context Length}\\\textbf{Violation}}} 
        & Asks the user for key information & \ding{51} & & & \ding{51} \\ \cdashline{2-6}[0.5pt/2pt]
        & Automatically ignores irrelevant long context & & & & \ding{51} \\ \cdashline{2-6}[0.5pt/2pt]
        & System architecture filters long context & & \ding{51} & & \\ \cdashline{2-6}[0.5pt/2pt]
        & No corrective behavior & & &  &  \\ 
        \midrule

        \multirow{6}{*}{\makecell[l]{\textbf{Hallucination}}} 
        & Infers true intent while executing misinformation & & & & \ding{51} \\ \cdashline{2-6}[0.5pt/2pt]
        & Fully accepts and executes misinformation & & & & \ding{55} \\ \cdashline{2-6}[0.5pt/2pt]
        & Partially accepts and executes misinformation & & & & \ding{55} \\ \cdashline{2-6}[0.5pt/2pt]
        & Ignores misinformation & \ding{51} & & & \ding{51} \\ \cdashline{2-6}[0.5pt/2pt]
        & Detects misinformation and seeks correct information & \ding{51} & & & \ding{51} \\ \cdashline{2-6}[0.5pt/2pt]
        & Uses external information sources to repair & \ding{51} & & & \ding{55} \\ \midrule

        \multirow{4}{*}{\makecell[l]{\textbf{Action Fault}}} 
        & Agent identifies and corrects  & \ding{51} & & & \ding{51} \\ \cdashline{2-6}[0.5pt/2pt]
        & Uses partial repair coordination or multipath compensation  & \ding{51} & & & \ding{51} \\ \cdashline{2-6}[0.5pt/2pt]
        & System uses redundancy or retry mechanisms & \ding{51} & & & \\ \cdashline{2-6}[0.5pt/2pt]
        & Responds to the error but ultimately fails & \ding{51} & & & \ding{55} \\ \cdashline{2-6}[0.5pt/2pt]
        & No corrective behavior  & & & & \\ 
        \midrule

        \multirow{7}{*}{\makecell[l]{\textbf{Role Ambiguity}}} 
        & Agent remains unaffected and follows original role & & & \ding{51} & \\ \cdashline{2-6}[0.5pt/2pt]
        & Agent disturbed with no system fault tolerance & & & \ding{55} & \\ \cdashline{2-6}[0.5pt/2pt]
        & Agent disturbed and system filters erroneous output & & \ding{51} & \ding{55} & \\ \cdashline{2-6}[0.5pt/2pt]
        & Agent disturbed and other agents compensate successfully & \ding{51} & & \ding{55} & \\ \cdashline{2-6}[0.5pt/2pt]
        & Agent disturbed and system fails to filter erroneous output & & \ding{55} & \ding{55} & \\ \cdashline{2-6}[0.5pt/2pt]
        & Agent disturbed and other agents fail to compensate & \ding{55} & & \ding{55} & \\ \cdashline{2-6}[0.5pt/2pt]
        & Agent disturbed but self compensates successfully & \ding{51} & & \ding{55} & \\ \midrule

        \multirow{5}{*}{\makecell[l]{\textbf{Blind Trust}}} 
        & Fully accepts incorrect information from other agents & & & \ding{55} & \\ \cdashline{2-6}[0.5pt/2pt]
        & Judges independently but correction fails & & & \ding{51} & \ding{55} \\ \cdashline{2-6}[0.5pt/2pt]
        & Judges independently and corrects successfully & & & \ding{51} & \ding{51} \\ \cdashline{2-6}[0.5pt/2pt]
        & Judges independently and other agents compensate & \ding{51} & & \ding{51} & \ding{51} \\ \cdashline{2-6}[0.5pt/2pt]
        & Makes no independent judgment & \ding{51} & & \ding{51} & \\ \midrule

        \multirow{7}{*}{\makecell[l]{\textbf{Instruction Logic}\\\textbf{Conflict}}} 
        & Reconciles conflicting instructions & & & & \ding{51} \\ \cdashline{2-6}[0.5pt/2pt]
        & Ignores some conflicts and rationalizes others & & & & \ding{51} \\ \cdashline{2-6}[0.5pt/2pt]
        & Ignores all conflicting instructions & & & & \ding{51} \\ \cdashline{2-6}[0.5pt/2pt]
        & Attempts to correct conflicting instructions & & & & \ding{51} \\ \cdashline{2-6}[0.5pt/2pt]
        & Detects conflicting instructions and requests clarification & \ding{51} & & & \ding{51} \\ \cdashline{2-6}[0.5pt/2pt]
        & Identifies conflicts but lacks a clarification mechanism & \ding{55} & & & \ding{51} \\ \cdashline{2-6}[0.5pt/2pt]
        & Fails to identify conflicting instructions & & & & \ding{55} \\ \midrule

        \multirow{4}{*}{\makecell[l]{\textbf{Instruction}\\\textbf{Ambiguity}}} 
        & Recognizes vague goal and guesses & & & & \ding{51} \\ \cdashline{2-6}[0.5pt/2pt]
        & Recognizes vague goal and asks user & \ding{51} & & & \ding{51} \\ \cdashline{2-6}[0.5pt/2pt]
        & Recognizes vagueness but architecture prevents asking & \ding{55} & & & \ding{51} \\ \cdashline{2-6}[0.5pt/2pt]
        & Fails to notice instruction ambiguity & & & & \ding{55} \\ \midrule

        \multirow{3}{*}{\makecell[l]{\textbf{Message Storm}}} 
        & System correctly filters messages & & \ding{51} & & \\ \cdashline{2-6}[0.5pt/2pt]
        & System fails to filter and duplicates do not affect behavior & & \ding{55} & & \ding{51} \\ \cdashline{2-6}[0.5pt/2pt]
        & System fails to filter and duplicates cause abnormal behavior & & \ding{55} & & \ding{55} \\ \midrule

        \multirow{3}{*}{\makecell[l]{\textbf{Message}\\\textbf{Cycle}}} 
        & System correctly filters messages & & \ding{51} & & \\ \cdashline{2-6}[0.5pt/2pt]
        & Receives cyclic messages without impact & & \ding{55} & & \ding{51} \\ \cdashline{2-6}[0.5pt/2pt]
        & Receives cyclic messages and behavior is affected & & \ding{55} & & \ding{55}  \\ \midrule

        \multirow{3}{*}{\makecell[l]{\textbf{Broadcast}\\\textbf{Amplification}}} 
        & System correctly filters messages & & \ding{51} & & \\ \cdashline{2-6}[0.5pt/2pt]
        & System fails to filter and irrelevant messages do not affect behavior & & \ding{55} & & \ding{51} \\ \cdashline{2-6}[0.5pt/2pt]
        & System fails to filter and irrelevant messages cause abnormal behavior & & \ding{55} & & \ding{55} \\ 
        \bottomrule
        
    \end{tabularx}
\end{table*}

\subsubsection{Behavioral Taxonomy and Fault Tolerance Dimensions}

To understand how MAS respond to and recover from faults, we analyze execution logs collected during fault injection. Table~\ref{tab:agent_behavior_tag} presents a comprehensive behavioral mapping across all 15 fault categories, cataloging the diverse fault-tolerant behaviors observed when agents encounter different types of failures.

The observed behaviors exhibit substantial heterogeneity rather than fixed failure patterns. For instance, under \textit{Instruction Ambiguity}, an agent may leverage its own reasoning to recognize missing details and proactively request clarification from the user. 
However, in the absence of a feedback channel (e.g., a tool that allows querying the user), the same situation may instead lead to brief confusion followed by autonomous continuation based on assumptions.
Under \textit{Message Storm}, redundant messages may be filtered out before reaching the agent, leaving its behavior unaffected; without such pre-filtering, the agent may still detect redundancy and ignore duplicated content on its own.
When facing \textit{Role Ambiguity}, an agent can resolve confusion by referring to examples provided by it's system prompt and realign itself with the intended responsibility. 
Even if prompt interference temporarily induces incorrect agent outputs, errors can be mitigated or overridden by other collaborating agents through cross-checking and compensation mechanisms.


These observations reveal that behaviors often involve multiple mechanisms: 
architectural structures (e.g., linear workflows), programmed logic (e.g., format validation), prompt design (e.g., role specification), and model-level reasoning (e.g., contextual inference).
However, simple behavior identification does not reveal \textit{which specific mechanisms} enable these behaviors or \textit{where resilience originates} within MAS architectures.
To address this gap, we derive four hierarchical fault tolerance tiers that classify behaviors by their source of resilience:



\begin{itemize}[noitemsep,leftmargin=5.5mm]
    \item \textbf{Mechanism-Level FT.} Fault tolerance derived from the system's structural design and temporal redundancy mechanisms. This includes architectural features such as iterative critique loops, multi-agent voting schemes, and redundant execution paths. These mechanisms operate independently of agent reasoning and are embedded in the MAS coordination infrastructure. 

    \item \textbf{Rule-Based FT.} Fault tolerance emerging from explicit procedural logic and heuristic rules encoded in the MAS implementation. This includes automatically deduplicates redundant messages. These behaviors are deterministic and activate when predefined conditions are met, regardless of the underlying model's reasoning capabilities.

    \item \textbf{Prompt-Level FT.} Rooted in the semantic robustness of User prompts. It leverages prompt engineering to guide agents through edge cases, clarify ambiguities, and maintain role boundaries, thereby pre-empting and mitigating faults.
    
    \item \textbf{Reasoning-Level FT.} Driven by the agent's high-level cognitive reflection. It relies on the underlying model's semantic understanding to autonomously detect logical inconsistencies, infer missing context, and resolve conflicts through multi-agent debate and consensus-building.
\end{itemize}
These four tiers often operate synergistically, and their interactions reveal the complexity of fault tolerance in MAS. In \textit{Blind Trust} scenarios, the behavior ``Judges independently but correction fails'' highlights a success in \textit{Prompt-Level FT}. While agents maintain their role definitions despite injected instructions, \textit{Reasoning-Level FT} fails because they cannot identify the erroneous nature of preceding inputs. Consequently, agents rationalize incorrect information instead of challenging it, demonstrating that role consistency does not guarantee semantic validation.
Similarly, in \textit{Instruction Logic Conflict} scenarios, the behavior ``Detects conflicts but architecture prevents querying'' demonstrates a case where \textit{Reasoning-Level FT} succeeds while \textit{Mechanism-Level FT} fails. Although agents identify logical inconsistencies in conflicting instructions, the absence of querying modules in the system architecture precludes recovery. Despite attempting to seek clarification to resolve ambiguity, agents are restricted by architectural constraints from accessing alternative information sources. This shows how \textit{Reasoning-Level FT} can identify semantic problems even when \textit{Mechanism-Level FT} fails to provide the infrastructure support necessary for recovery actions.

\subsubsection{Empirical Evaluation of Fault Tolerance Tiers}

\begin{figure*}[tp] 
    \centering
    
    \begin{subfigure}{\linewidth}
        \centering
        \includegraphics[width=0.9\linewidth]{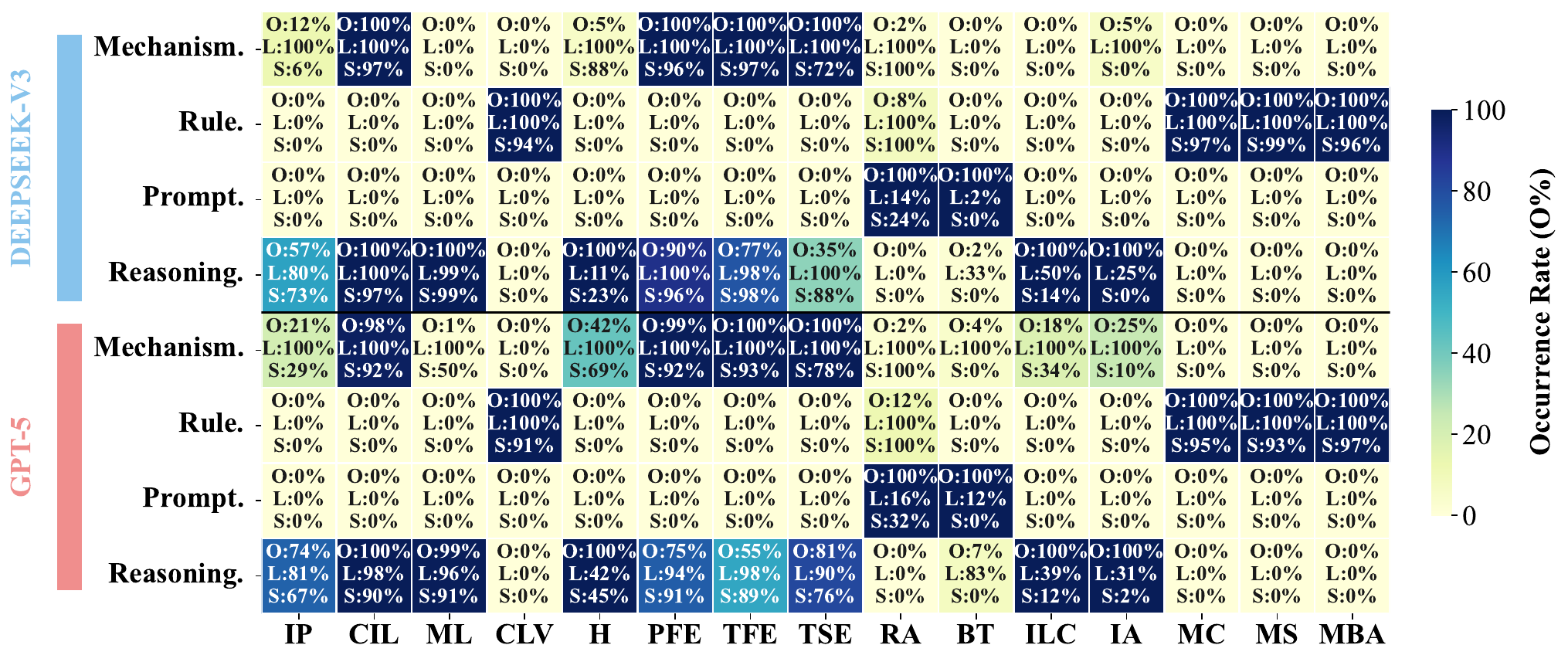}
        \caption{MetaGPT}
    \end{subfigure}
    \vfill 
    
    \begin{subfigure}{\linewidth}
        \centering
        \includegraphics[width=0.9\linewidth]{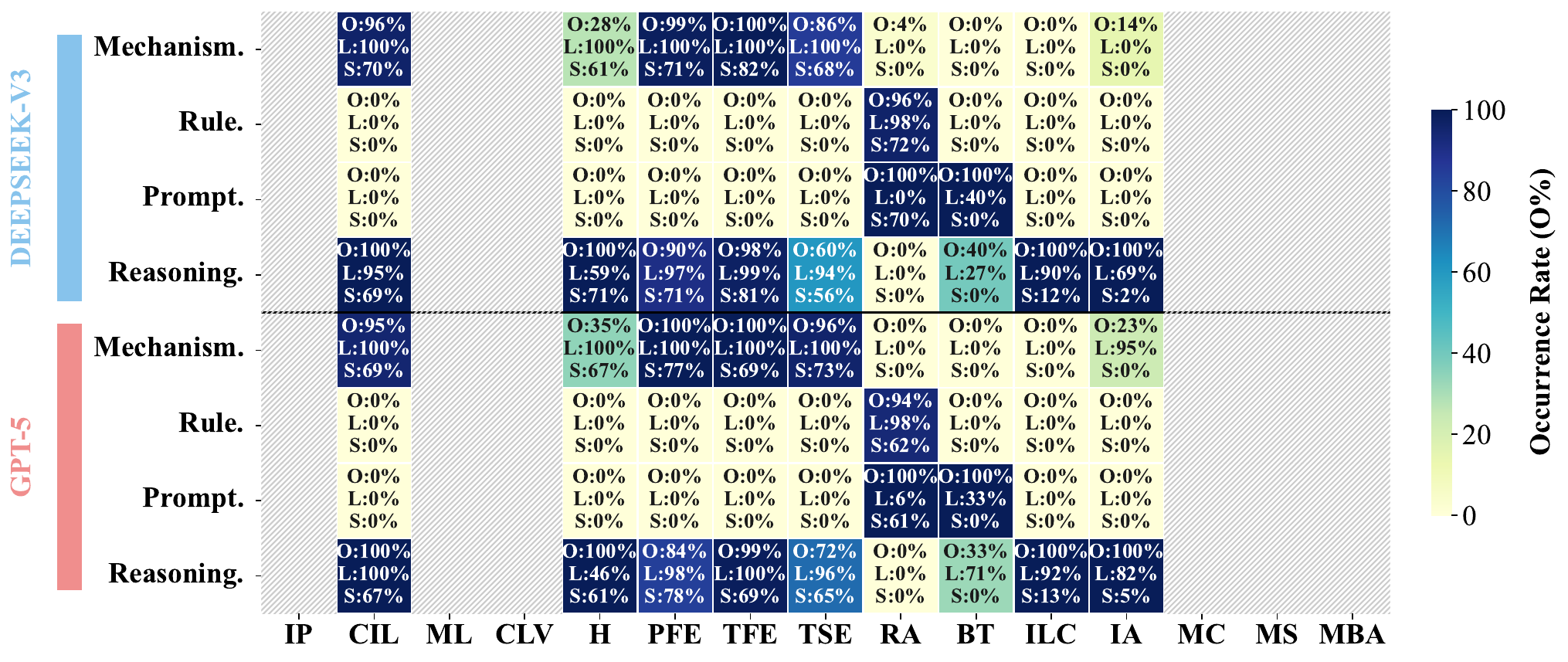}
        \caption{Camel}
    \end{subfigure}
    \vfill
    
    \begin{subfigure}{\linewidth}
        \centering
        \includegraphics[width=0.9\linewidth]{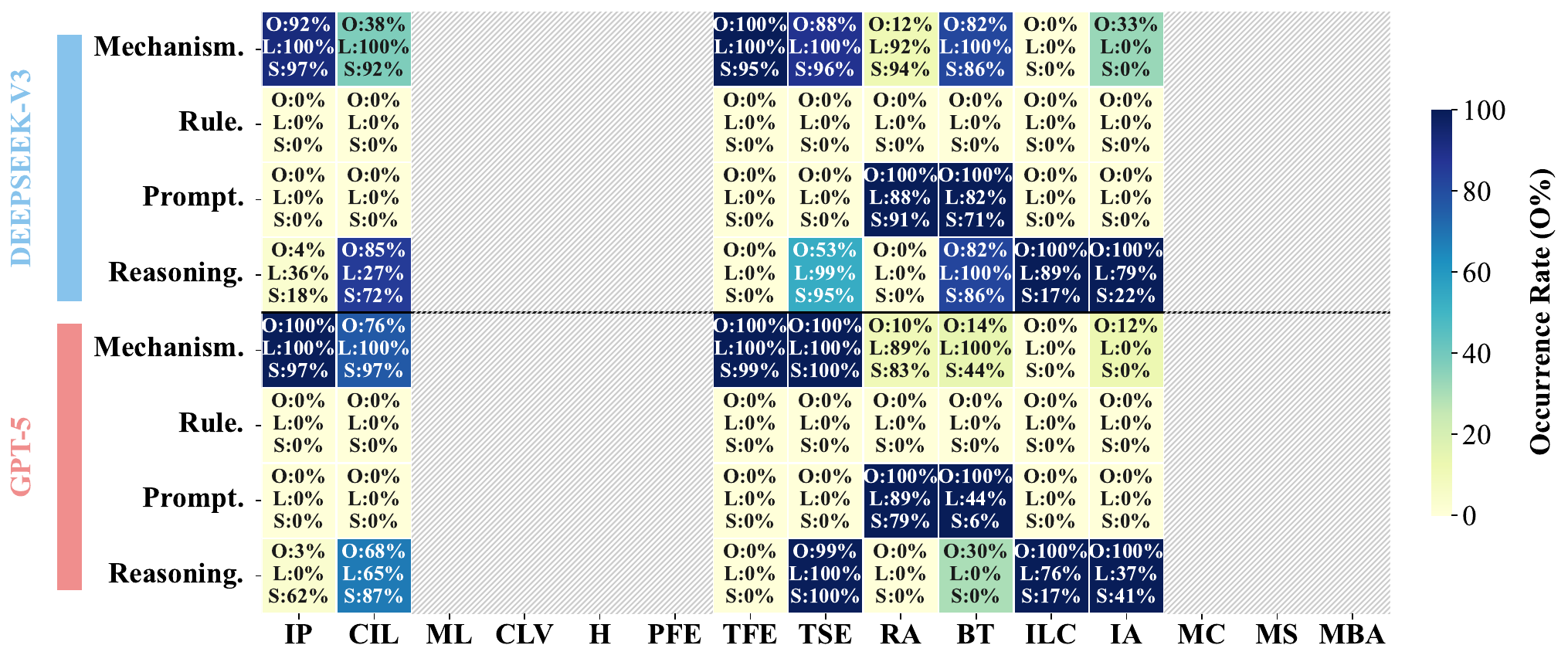}
        \caption{Table-Critic}
    \end{subfigure}

    \caption{Fault-tolerance Performance of Different MAS under 15 Fault Types. Gray columns indicate that the corresponding faults cannot be injected due to system architecture limitations or output format constraints.}
    \label{fig:tag_results}
\end{figure*}

Our analysis of fault-tolerance performance across the four hierarchical tiers reveals distinct characteristics for each layer. The fault-tolerance performance across the four tiers is visualized in the heatmap of Fig. \ref{fig:tag_results}, evaluated through the $(O_f, L_f, S_f)$ triplet defined in Sec.~\ref{ssec:processlevel}. 

Under perturbations including \textit{Parameter Filling Error}, \textit{Tool Format Error}, and \textit{Tool Selection Error}, the systems demonstrate robust Mechanism-Level fault tolerance. Across all systems, the occurrence rates consistently reach $O_f \ge 85\%$, local recovery rates maintain $L_f = 100\%$, and final task success rates are secured at $S_f > 61\%$.

\begin{mdframed}
\textbf{Finding 6:}
Mechanism-Level defenses (e.g., automated retries, syntax parsers) function as a filter for low-level errors within MAS. By effectively handling execution-layer noise, these mechanisms prevent low-level errors from propagating into the cognitive context of agents, ensuring the accuracy of historical information within high-level reasoning processes.
\end{mdframed}

\textit{Rule-Based FT} emerges from explicit procedural logic and deterministic rules hardcoded in the MAS implementation. Unlike \textit{Mechanism-Level FT} which relies on architectural redundancy and retry mechanisms, or \textit{Reasoning-Level FT} which depends on semantic understanding, \textit{Rule-Based FT} operates through programmed exception handling that detects and filters predefined structural patterns. This tier proves particularly effective for structural anomalies such as \textit{Communication Faults} (message storms, cycles, broadcast amplification) and \textit{Context Length Violation}, where Mechanism and Reasoning layers fail because these faults require deterministic pattern matching rather than semantic interpretation or architectural compensation. In MetaGPT, \textit{Rule-Based FT} implemented as hardcoded filtering rules achieves perfect detection and recovery: $O=100\%$ (all fault instances trigger the filter) and $L=100\%$ (all detected faults are successfully resolved), resulting in task success rate $S > 93\%$. This demonstrates that deterministic procedural logic provides guaranteed mitigation for well-defined structural patterns, i.e., once activated, recovery is certain.

\begin{mdframed}
\textbf{Finding 7:} \textit{Rule-Based FT} provides deterministic handling of structural failures through programmed exception handling. In MetaGPT, \textit{Rule-Based FT} achieves $100\%$ recovery rates for communication-related anomalies and context overflows, indicating that procedural logic stabilizes systems when communication protocols or memory constraints are exceeded.
\end{mdframed}

Prompt-Level fault tolerance relies on the semantic robustness of agent instructions encoded in system and user prompts. In contrast to \textit{Rule-Based FT} which operates through hardcoded rules, \textit{Prompt-Level FT} depends on agents' ability to interpret and adhere to textual directives that define roles, responsibilities, and behavioral constraints. For \textbf{Configuration Faults} such as \textit{Role Ambiguity} and \textit{Blind Trust}, \textit{Prompt-Level FT} achieves universal activation ($O_f=100\%$) across all architectures, as these faults directly modify system prompts and invariably trigger Prompt-Level responses. However, activation does not guarantee successful recovery. The effectiveness depends on whether corrupted prompts preserve or undermine the semantic foundations necessary for correct agent behavior.
The recovery success rates ($S_f$) diverge sharply across fault types and architectures. For \textit{Role Ambiguity}, Table-Critic exhibits high resilience ($S_f=91\%$ for DeepSeek-V3, $S_f=79\%$ for GPT-5) as corrupted role definitions still allow agents to retain core task-solving capabilities, while MetaGPT achieves lower rates ($S_f \in [24\%, 32\%]$) and Camel maintains intermediate performance ($S_f \in [61\%, 70\%]$). Conversely, \textit{Blind Trust} induces systemic collapse in Camel ($S_f=0.0\%$) and MetaGPT ($S_f=0.0\%$) as corrupted prompts instruct agents to unconditionally accept erroneous upstream information, fundamentally undermining their verification capabilities. Only Table-Critic shows partial resilience ($S_f=71\%$ for DeepSeek-V3, $S_f=6.32\%$ for GPT-5) by leveraging Reasoning and Mechanism layers to override compromised directives.

\begin{mdframed}
\textbf{Finding 8:}
Prompt modifications universally trigger FT ($O_f=100\%$), but efficacy is fault-dependent. In \textit{Role Ambiguity}, success rates vary by architecture (Table-Critic: $S_f \in [79\%, 91\%]$; MetaGPT: $S_f \in [24\%, 32\%]$) as agents retain task-solving logic. In \textit{Blind Trust}, Prompt-Level defenses fail in MetaGPT and Camel ($S_f=0.0\%$) due to strict adherence to erroneous instructions, while Table-Critic shows partial resilience ($S_f \in [6\%, 71\%]$). Higher resilience requires Reasoning or Mechanism-layer interventions to override corrupted directives.
\end{mdframed}

As shown in Fig.~\ref{fig:tag_results}, the highest tier of fault tolerance (\textit{Reasoning-Level FT}) addresses semantic faults that bypass lower-tier defenses. Three fault types (\textit{Hallucination}, \textit{Instruction Logic Conflict}, \textit{Instruction Ambiguity}) share a common profile, i.e., syntactically valid but semantically defective, allowing them to evade Mechanism-Level and Rule-Based filters that rely on structural pattern matching. Mechanism-Level occurrence rates ($O_f$) vary dramatically across these faults. For \textit{Hallucination}, MetaGPT achieves low detection ($O_f=5.5\%$ for DeepSeek-V3, $O_f=42.2\%$ for GPT-5) as hallucinations originate from internal reasoning, while Camel reaches higher rates ($O_f \in [27.7\%, 34.7\%]$) because external information sources enable partial detection. For \textit{Instruction Logic Conflict}, Mechanism-Level detection is nearly absent ($O_f=0.0\%$ for most systems; only MetaGPT with GPT-5 reaches $O_f=18\%$). For \textit{Instruction Ambiguity}, detection remains limited (MetaGPT: $O_f \in [5.5\%, 24.8\%]$; Table-Critic: $O_f \in [12.4\%, 32.9\%]$; Camel: $O_f \in [14.5\%, 23.2\%]$). In stark contrast, \textit{Reasoning-Level FT} achieves universal activation ($O_f=100\%$) across all three fault types and all systems, serving as the primary and often sole defense against semantic anomalies.



\begin{mdframed}
\textbf{Finding 9:}
Faults like \textit{Hallucination}, \textit{Logic Conflict}, and \textit{Ambiguity} are syntactically correct but semantically defective, allowing them to bypass Mechanism-Level filters. \textit{Reasoning-Level FT} is the primary defense ($O_f=100\%$), where agents leverage cognitive redundancy to detect semantic flaws and infer correct intent to resolve instruction-level errors.
\end{mdframed}

\section{Discussion} 

\subsection{Treat Upstream Instructions with Caution}

Strict adherence to system prompts and upstream instructions is widely regarded as desirable in MAS. However, our evaluation reveals that this assumption breaks down under fault conditions. When instructions are corrupted or internally inconsistent (Sec.~\ref{subsec:rq1}), rigid compliance becomes a liability. Under \textit{Blind Trust}, GPT-5's superior instruction-following led to near-total collapse ($RS_f = 6.32\%$), while DeepSeek-V3's weaker compliance paradoxically preserved functionality ($RS_f = 70.61\%$).

These results expose a fundamental design tension: agents must be compliant enough to follow valid directives yet skeptical enough to detect corrupted ones. Robust MAS should incorporate \textit{conditional compliance} mechanisms, allowing agents to pause execution or flag inconsistencies when they encounter logical contradictions, constraint violations, or conflicting environmental feedback. Rather than treating all upstream signals as authoritative, agents should cross-validate instructions against their own reasoning before committing to irreversible actions, preventing locally corrupted directives from cascading into system-level failures.


\subsection{Avoid Failure Propagation in Linear Agent Workflow}
Linear, pipeline-style workflows are widely adopted in MAS due to their simplicity and clear stage boundaries. However, our results show this topology is the most vulnerable to cascading failures. Under Configuration and Instruction Faults, MetaGPT's linear pipeline collapsed to $RS_f$ as low as $0.0\%$, as a single corrupted output propagates downstream unchecked, with each subsequent agent inheriting and compounding the error. In contrast, Table-Critic's iterative closed-loop maintained significantly higher robustness by enabling repeated validation and correction cycles.

The core vulnerability is single-path dependency: no redundancy exists to catch semantic drift before it reaches downstream consumers. We identify two complementary mitigation strategies. First, \textit{multi-source validation}: downstream agents should reconcile information from multiple independent sources, such as parallel agent interpretations or environmental ground truth, to detect inconsistencies before acting. Second, \textit{inline verification checkpoints}: lightweight validation stages between pipeline steps can assess output plausibility (e.g., schema conformance, semantic consistency with the task specification) and trigger re-execution when anomalies are detected. These mechanisms introduce the error-correction benefits of closed-loop architectures while preserving the interpretability of linear workflows.


\section{Related Work}

\subsection{Fault Injection and Chaos Engineering}
Traditional fault injection and mutation testing focus on low-level syntactic corruptions such as memory leaks~\cite{hsueh1997fault}. These methodologies are ill-equipped for Multi-Agent Systems (MAS) where failures manifest as semantic deviations. Unlike deterministic faults, semantic failures allow a system to remain operational while becoming logically decoupled from its intended tasks. Current chaos engineering and static metrics fail to capture the dynamic nature of these agentic collapses~\cite{andries2024}. For example, a system may reach a superficial consensus even when its internal reasoning has drifted into a hallucinated state, producing outputs that are superficially coherent yet logically invalid.

\subsection{MAS Evaluation Frameworks}
Recent studies explore how localized failures propagate across MAS topologies. AutoInject~\cite{faulty-agents-resilience} demonstrates that fault propagation patterns depend critically on underlying organizational structures. Despite the importance of organizational interactions, existing MAS evaluation frameworks remain limited in systematically diagnosing interaction-level failures. While adversarial benchmarks like TAMAS~\cite{kavathekar2025tamas} address intentional sabotage, the more pervasive threat in production remains spontaneous coordination failure.  Current frameworks rely on coarse-grained outcome metrics such as task success rates or binary pass/fail. While frameworks like AppWorld~\cite{AppWorld}, AgentBoard~\cite{agentboard}, and ScienceAgentBench~\cite{ScienceAgentBench} incorporate sub-goal tracking to move beyond binary success, they remain outcome-oriented. Their metrics focus on linear task progress rather than interaction-layer resilience. OpenJudge~\cite{openjudge} offers multi-dimensional monitoring capabilities but relies on idealized scenarios, lacking the stress tests to evaluate organizational resilience. 

\section{Conclusion}

This paper introduced \alias, a framework designed to diagnose and evaluate the robustness of Multi-Agent Systems through systematic fault injection. Through the lens of 15 distinct fault types, we have shown that MAS reliability depends on a complex interplay between foundation model reasoning and coordination infrastructure. Our findings challenge the assumption that model scaling alone ensures system stability. Instead, we quantified the superior protective power of specific architectural patterns, such as shared message pools and iterative critique loops, which effectively neutralize semantic errors before they propagate into systemic collapse. By providing a granular behavioral taxonomy and a suite of process-oriented metrics, \alias equips the software engineering community with a rigorous methodology to evaluate, diagnose, and harden the new generation of orchestrated intelligent software.

\section*{Data Availability}
The experimental data of \alias are publicly available on \url{https://github.com/wxhhxn/MASFIRE}.


\bibliographystyle{ACM-Reference-Format}
\bibliography{references}


\end{document}